\def\firstpage{1}
\def\lastpage{14}
\def\mmamid{XXXX.xxxxxx}
\def\firstfootnote{0}
\def\mmatitle{Pricing of Warrants with Stock Price Dependent Threshold Conditions}
\def\mmarunauthor{A. Olvik \and R. Kangro}
\def\mmaruntitle{Warrants with Threshold Conditions}
\def\mmakeywords{warrants, threshold conditions, dilution effect, indifference pricing}
\def\mmaams{91G20}
\def\mmareceived{\today}
\def\mmarevised{\today}

\documentclass{mma}
 \MMAJCLS
 \makeatletter
 \spnewtheorem*{west}{Test the West}{\bfseries}{\itshape}

 \@ifundefined{BibTeX}{%
 }{}
 \makeatother
  
    \usepackage{tikz}		
    \usetikzlibrary{matrix} %
    \usepackage{graphicx}	

\begin{document}
\bibliographystyle{plainmma}
\begin{topmatter}
%
  \title{\mmatitle\thanks{This research has been supported by Estonian Research Council under ETF grant nr 8802, Estonian Ministry of Education and Research under projects IUT34-5 and SF0180015s12 and by Estonian Doctoral School in Mathematics and Statistics.} }
 \author{Ander Olvik \and Raul Kangro}
\vskip 0.2cm
 \institution{Institute of Mathematical Statistics, University of Tartu}
 \address{{\ }J.Liivi 2, 50409 Tartu, Estonia}
  \Email{ander1@ut.ee}
 \end{topmatter}
\MMAJarticle
 \begin{abstract}
Warrants with stock price dependent threshold conditions give the right to buy specially issued stocks, if the performance of the stock price satisfies some requirements. Existence of these derivatives changes the price process of the underlying. We show that in the presence of such warrants one cannot assume that the stock market is arbitrage free and that the stock is tradeable at every time moment with the same price for buying and selling. This means that the usual methods for deriving fair prices for such warrants cannot be used.  We start from a simple model for the firm's value process and discuss some ways to specify a related model for the stock price process in the presence of warrants with threshold conditions. We also discuss how indifference pricing approach can be used for pricing such warrants.
 \end{abstract}
 \Keywords 
%

\section{Introduction}\label{s:1}

A traditional (or corporate) warrant is a derivative that gives its owner the right to buy the underlying stock at a fixed time for a fixed price. It differs from a call option contract since the issuer is the company of the underlying and upon exercising warrants, new shares are issued. Since there are more shares, the value of each share decreases. This is known as the dilution effect\footnote{Covered (or naked) warrants do not cause dilution and are not examined in this paper.}. These derivatives are characterized by long lifetimes and low or non-existing trading activity on the aftermarket. This makes it difficult to estimate the fair value by just marking them to the market. Such derivatives are suitable for motivating employees and may occur in the literature also by the name of employee stock options\footnote{Which, in this case, also issue new stocks when exercised.} (ESOs).

In this paper we examine warrants with stock price dependent threshold conditions. Such warrants are not very common but have been used in practise (see, for example, the 2010 annular report of Trigon Capital Group, \cite[p.~44]{Trigon}). We consider European type warrants that can be exercised only if during the lifetime of the warrant the underlying stock price reaches a predetermined threshold level.  We use the term \textbf{threshold warrant}, which stresses two main characteristics of the derivative contracts in question. Firstly, like warrants  (and unlike traditional options) they exhibit dilution effect. Secondly, (unlike typical warrants and ESOs) exercising rights are dependent on the performance of the underlying. 

Like with every other derivative security, in order to price the warrants, we need to model the underlying. For pricing classical warrants, it is sufficient to model the value of the company. Usual assumptions similar to the ones made in the option pricing theory allow the derivation of the equation (and for simple models also the formula) for warrant prices in terms of the value of the productive assets of the company (see \cite{GalaiS1978, Hull}).   However, in the case of threshold warrants, we specifically need to know whether the stock price satisfies the exercise conditions. As the exercising of warrants increases both the number of shares and brings extra money to the firm's balance sheet, the relationship between firm's value and stock price is not linear any more. For example, the log-normality of company's value does not imply log-normality of stock price \cite{GalaiS1978}. A simple solution would be to assume that out of the two, it is the stock price that follows some commonly used stochastic process, e.g. geometric Brownian motion\footnote{This might be a practical approach in case of complicated exercise conditions. Then the price estimate is dominated by our ability to simulate in-the-money trajectories. Also, computing time of the MC method can be significantly reduced with such assumption.} and ignore company's value process that remains unknown. 
Such approach contradicts the viewpoint that the issuance of warrants should not affect the value process of the productive assets of a company. For example, if a geometric Brownian motion is used for modelling the stock price before the issuance of warrants like in \cite{BS1973} and thus the value of the company also corresponds to a geometric Brownian motion, we should use the geometric Brownian motion to model the value of the assets of the company during the lifetime of the warrants. Then it is the stock price process that is  log-normal before the warrants are issued, but changes after the warrants' issuance.

It is quite well known how the dilution effect of traditional warrants impacts the behaviour of the stock price process and how observable information about stock prices can be used for determining the value of the warrants \cite{Ingersoll77,Daves_Ehrhardt, LiWong, Ukhov, Galai89}. As shown in \cite{HankeP2002}, the knowledge about the behaviour of the stock price process in the presence of warrants is also necessary for the consistent pricing of any outstanding common options. As far as we know, the impact of the presence of warrants with stock price dependent threshold conditions on the stock price process has not been studied before. 

In this paper we show that in the presence of threshold warrants one cannot have an arbitrage free stock price model that is consistent with the stock price model in the presence of traditional warrants after the threshold condition is met. This means that the usual methods for deriving fair prices for such warrants cannot be used and some of the usual assumptions about the behaviour of the stock market have to be relaxed in order to compute prices of threshold warrants.

The paper is organized as follows: section 2 introduces the notation and the framework for our analysis. In section 3 we demonstrate that the presence of threshold warrants introduces arbitrage opportunities. In section 4 we discuss the need to relax common assumptions of arbitrage fee pricing models for the consistent pricing of threshold warrants. Section 5 introduces indifference pricing approach. Section 6 provides a summary and directions for further research.

\vspace{1cm}
\section{Notation and framework}\label{s:2}
Throughout the paper we will use the following notation:
\begin{tabbing}
$X(t)$ ~ \= - value of the company at time $t$;\\ 
$T$ \> - time to maturity;\\
$K$ \> - exercising price;  \\
$L$ \> - threshold price level;\\ 
$N$ \> - number of shares before warrants are exercised;\\
$M$ \> - number of warrants.\\
\end{tabbing}

By $X(t)$ we denote the value of a company's productive business assets. If we assume an efficient market model (which we silently do, when we assume that prices follow random walk model \cite{Walter03theefficient}), then it can be interpreted as the market value of equity that could be observed (by multiplying stock price with number of shares) unless the company had issued warrants at time $t=0$. Later on it becomes unobservable without knowing the behaviour of stock and warrant prices - in order to buy the whole company one needs to buy every existing stock and warrant.

We denote two types of fractions of firm value. The first one of them corresponds to a stock price in a situation, where warrants will certainly not be exercised (or do not exist) and the other to a case where warrants will certainly be exercised:
\begin{equation}
S(t) = \frac{X(t)}{N}, \qquad S'(t) = \frac{X(t)+ D_{t,T}\!\cdot\! MK}{N+M},
\end{equation}
where $D_{t,T}$ is the discount factor applicable to the time period from $t$ to $T$, $0 < D_{t,T} \leq 1$, $D_{T,t} = 1/D_{t,T} $.
By stock price we mean the price at which a share is traded. In presence of $M$ traditional warrants, it is denoted $S_{W}(t)$. In presence of $M$ threshold warrants, it is denoted $S_{W_{L}}(t)$. Warrants may be exercised if $S_{W_{L}}(t) \geq L$ for some $t \in (0,T)$. Each warrant grants the right to buy one share for price $K$. Threshold $L > S_{W_{L}}(0)$ and $L>K$\footnote{Otherwise the threshold would essentially not exist.}. \\

Assumptions:\\
\textbf{A1.} \ Proceeds from warrants' issuance do not affect firm value process\footnote{They are distributed as dividends or provide the same return as the rest of the firm's assets.};\\
\textbf{A2.} \ Warrant holders cannot affect the market in favourable direction;\\
\textbf{A3.} \ No other outstanding equity securities;\\
\textbf{A4.} \ No information asymmetry\footnote{It also means that all market participants are aware of the existence of warrants.};\\	
\textbf{A5.} \ No transaction costs;\\
\textbf{A6.} \ No dividends.\\

All in all, in current setup the firm's value $X(t)$ is the only 
source of randomness.
It drives the value of $S_{W_{L}}(t)$, which in turn determines the warrant price.
\vspace{1cm}

\section{Properties of the stock price process after warrants issuance}\label{s:3}
During the lifetime of a warrant, the stock price process differs before and after reaching the threshold. First we examine the latter case.

\subsection{Stock price process after reaching exercising threshold}

If the exercising condition is met, then the threshold warrant is equal to a classical warrant. The stock price process changes accordingly\footnote{In discrete time model, if $S_{W_{L}}(t) \geq L$, then $S_{W_{L}}(t\!+\!1) = S_{W}(t\!+\!1)$.}. 
This knowledge is useful, when one is simulating price trajectories or estimating market parameters. Exercising of classical warrants is only subject to profitability. Exercising is profitable, if $S'(T)>K$\footnote{The profitability requirement can be written as 
\[S'(T)>K \quad \Longleftrightarrow \quad \frac{X(T)+MK}{N+M}-K>0 \quad \Longleftrightarrow \quad X(T)>NK.\]}. 
The payoff function:
\begin{equation}
\textrm{max}\left( \frac{X(T)+MK}{N+M} - K, \: 0 \right) = \dfrac{N}{N+M} \cdot \textrm{max}\left( \dfrac{X(T)}{N} - K, \: 0 \right).
\end{equation}
By modelling $X(t)$ we obtain the well-known result (e.g. Galai and Schneller 1978) that the price of a warrant is a $N/(N+M)$ fraction of a call option of a similar firm but without warrants. This way we can work around the need to model the changed stock price process, by simply modelling the company's value $X(t)$. \\ 

For modelling $X(t)$, however, we need to estimate market parameters. Only at time $t=0$ has the firm value been previously directly observable. Later we need to know the relationship between $S_{W_{L}}(t)$ and $X(t)$. 
The general derivation of stock price process after warrants' issuance is given in \cite{HankeP2002}. Simply put, if the firm value $X(t)$ includes proceeding from warrants' issuance, then during the lifetime of warrants:\\
\begin{equation}
S_{W}(t) = \frac{X(t)}{N} - \frac{MW(t)}{N},
\end{equation}\\
where $S_{W}(t)$ denotes stock price in the presence of $M$ classical warrants and $W(t)$ is the price of one warrant. Methods for estimating unobservable $X(t)$ volatility (by using observable stock return variance) are described in \cite{Ukhov} and \cite{Daves_Ehrhardt}.
\vspace{1cm}

\subsection{Stock price process before reaching exercising threshold} 

\begin{figure}
\begin{center}
\includegraphics[scale=0.5, trim = 20mm 30mm 0mm 25mm, clip]{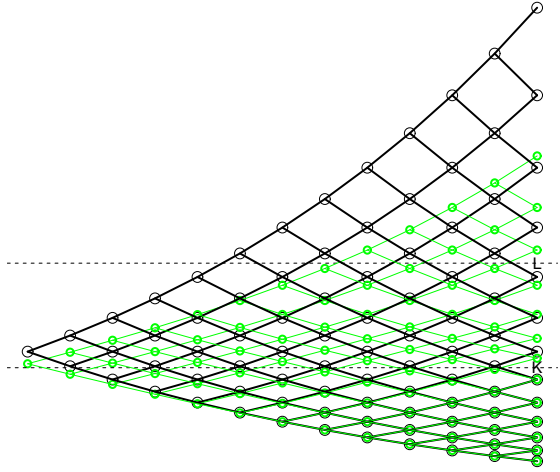} 
\caption{A tree of $S(t)$ (black) and $S_{W}(t)$ (green).}  
\end{center}
\end{figure}

The derivation of stock price model is complicated by the threshold condition, because the exercising condition requires the stock price to rise above a pre-set level. We show that even when assuming the firm value is known, in some situations the stock price is still unknown. Now we have two exercising  conditions:
\begin{itemize}
\item \	There exists a $t \in (0,T)$ at which the stock has been traded for the price $S_{W_{L}}(t)$ satisfying the threshold condition $ S_{W_{L}}(t) \geq L$;
\item \	Exercising is profitable, i.e. $S'(T)>K$.
\end{itemize}

\subsubsection{Existence of arbitrage opportunities}

In this section we illustrate our arguments by using a discrete time market model. Firm value and stock price processes are modelled by binomial trees as described by Cox et al. in \cite{CoxRR1979}. We will distinguish a specific node by notation $X(\tau,t)$, where $-T \leq \tau \leq T$ is the number of upticks from initial level by time $t$. We omit it, if we examine an arbitrary node at time $t$. An example of $S(t)$ and $S_{W}(t)$ trees is given in figure 1, where $X(0)=1000$, $N=10$, $M=7$, $L=155$, $K=90$ and uptick return ratio $U=1.1$.\\

Let's see a 2-period model. The profitability condition is examined at $t=T=2$, the threshold condition is examined only at $t=1$. 
As an example, in figure 2 we have trees of $S(t)$ and $S_{W}(t)$, where $X(0)=1000$, $N=10$, $M=3$, $L=108$, $K=95 < S(0)$, uptick return ratio $U=1.1$. At maturity
\[
 S_{W_{L}}(T) = 
  \begin{cases}
   S(T) = X(T)/N, &  S_{W_{L}}(1) < L, \\  
   S_{W}(T) = S'(T), &  S_{W_{L}}(1) \geq L.
  \end{cases}
\]

\begin{figure}
\begin{center}
\includegraphics[scale=0.5, trim = 20mm 30mm 0mm 20mm, clip]{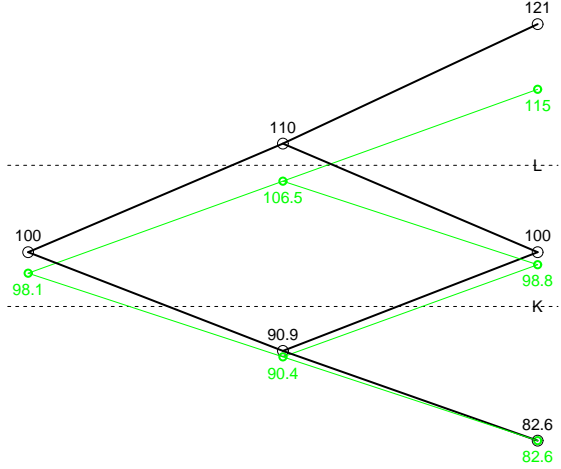} 
\caption{Trees of $S(t)$ (black) and $S_{W}(t)$ (green).} 
\end{center}
\end{figure}

If we assume that both the stock and the productive assets of the firm are tradeable, then according to the Fundamental Theorem of Asset pricing (see, for example, \cite{Pascucci}) the market consisting of $S(t)$ and $X(t)$ is arbitrage free if and only if there exists an equivalent probability measure (the so-called risk neutral measure) under which both the discounted stock price and the discounted firm value process are martingales. For the simplicity of the arguments, let us assume that the risk free interest rate is 0 ($D_{t,T}=1$). Then both $X(t)$ and $S(t)$ should be martingales under a risk neutral measure.
 
If we derive the risk neutral probabilities from the tree, then in our example's upper branch
\[
E( S_{W_{L}}(T) \! \mid \! X(+1,1)) = 
  \begin{cases}
   110  , &  S_{W_{L}}(+1,1) < 108, \\ 
   106.5, &  S_{W_{L}}(t)(+1,1) \geq 108.
   \end{cases}
\]
Therefore $E( S_{W_{L}}(t\!+\!1) \! \mid \! \mathcal{F}_t) \neq S_{W_{L}}(t)$ and \textbf{$S_{W_{L}}(t)$ is not a martingale!} \\

\vspace{0.5cm}
Next we will show, without specifying a probability measure, that $S_{W_{L}}(t)$ cannot be a martingale. For that we establish a possible region for stock price value. Note that
\[
S_{W_{L}}(t) \leq S(t).
\]
The inequality holds, since under existence of warrants, one stock may represent less than $1/N$-th of the equity. 
On the other hand, all firm value trajectories that lead to the exercising of threshold warrants belong to a set of trajectories which lead to the exercising of similar, but classical warrants. Therefore, 
due to a lesser chance of dilution realizing, we should have
\[
S_{W_{L}}(t) \geq S_{W}(t).
\]
Therefore, 
\begin{equation}
S_{W_{L}}(t) \in \left[S_{W}(t), \ S(t)\right].
\end{equation}
\vspace{0.1cm}

\begin{figure}
\begin{center}
\includegraphics[scale=0.5, trim = 20mm 30mm 0mm 25mm, clip]{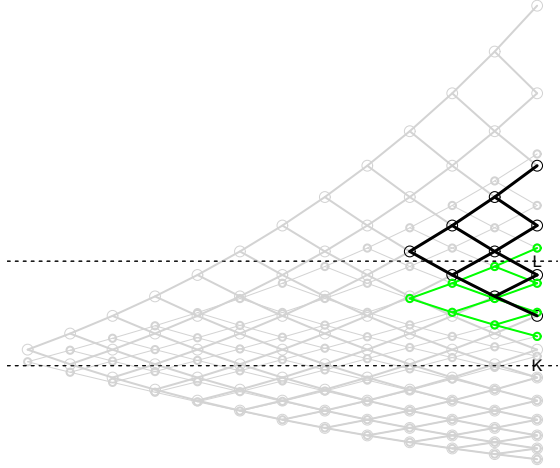} 
\caption{Highlighted subtrees of $S(t)$ (black) and $S_{W}(t)$ (green) from figure 1, which demonstrate that $S_{W_{L}}(t)$ is not a martingale.}  
\end{center}
\end{figure}

Let $m$ be the minimum number of upticks for
the undiluted share 
price to (theoretically) exceed the threshold $L$. With sufficient number of discrete time steps,
\begin{equation}
S(m, t) > L > S_{W}(m, t).
\end{equation}

Let us
take a closer look at the sections of $S(t)$ and $S_{W}(t)$ trees near the maturity date that correspond to the same firm values. We examine such trajectories, where $S(t)$ has not reached the threshold before our examined section but is close enough to $L$ and the number of discrete time steps is large enough (i.e. $K$ is low and $T$ close enough) so that upon achieving exercising rights, the profitable exercise will certainly occur. It means that in the subtree of $S_{W}(t)$ every node can be written as 
\begin{equation}
S_{W}(\tau, t) = S'(\tau, t) = ( X(\tau, t) + D_{t,T} \!\cdot\! MK ) / (N+M). 
\end{equation}
In figure 3, we have highlighted such subtrees from the trees in figure 1. Let time $t$ be the first timepoint, where in our subtree $S(\tau, t) = S(m, t) > L$. If the parameters $N$, $M$ and $K$ are such that,

\begin{equation}
S(m\!-\!1 ,t\!+\!1) = \frac{X_{0} \! \cdot \! U^{m} \! \cdot \! D }{N} > \frac{X_{0} \! \cdot \! U^{m} \! \cdot \! U + D_{t+1,T} \!\cdot\! MK}{N+M} = S_{W}(m\!+\!1, t\!+\!1),
\end{equation}

and transaction price exceeds threshold at $t$, then

\[
S_{W_{L}}(m, t) > L \overset{m \ \text{def.}}{>} S(m\!-\!1, t\!+\!1) \overset{\text{(3.6)}}{>} S_{W}(m\!+\!1, t\!+\!1),
\]

but if $S_{W_{L}}(t)$ were a martingale then $S_{W_{L}}(\tau, t) \leq S_{W_{L}}(\tau\!+\!1, t\!+\!1) = S_{W}(\tau\!+\!1, t\!+\!1)$\footnote{Recall that after attaining exercising rights, stock price should correspond to the existence of traditional warrant.}. So if the transaction price exceeds the threshold, then the stock price cannot be a martingale. Should $S_{W_{L}}(m,t)$ not exceed the threshold, we will examine the next upticks. Assuming sufficient discretization, large enough $t$ and large enough $\frac{M}{N}$, for all $m'\in \{0,1,\ldots,T-t-1\}$ upticks, following result holds:
\begin{equation}
S(m\!+\!m', t\!+\!m') > L > S_{W}(m\!+\!m', t\!+\!m').
\end{equation}
By multiplying both sides of inequality (3.6) by $U^{m'}$ we get (recall $D_{t,T}\!=\!1$):
\[
\frac{X_{0} \! \cdot \! U^{m\! + \! m' \! - \! 1}}{N} > \frac{X_{0} \! \cdot \! U^{m \!+\!m'\!+\!1} +  MK \! \cdot \! U^{m'}}{N+M} \overset{U>1}{>} \frac{X_{0} \! \cdot \! U^{m \!+\!m'\!+\!1} + MK}{N+M}, 
\]
\[
\textrm{i.e.} \quad S(m\!+\!m'\!-\!1, t \!+\!m'\!+\!1) > S_{W}(m\!+\!m'\!+\!1, t \!+\!m'\!+\!1).
\]
So, if after $m'$ upticks stock price would exceed $L$ and 
\begin{equation}
L > S_{W}(m\!+\!m'\!+\!1, t\!+\!m'\!+\!1)
\end{equation}
holds, it still means that stock price cannot be a martingale.\\

Let $m'$ be such that $t\!+\!m'\!+\!1 = T$. We saw that $S_{W_{L}}(m\!+\!m', T\!-\!1) \geq L$ means $S_{W_{L}}(t)$ cannot be a martingale but we reach the same conclusion if $S_{W_{L}}(m\!+\!m', T\!-\!1) < L$. In that case the exercising condition is not met and $S_{W_{L}}(T)=S(T)$, but assuming 
\begin{equation}
m' \geq 1, 
\end{equation}
we get
\[
S_{W_{L}}(m\!+\!m'\!-\!1, T) = S(m\!+\!m'\!-\!1, T) \overset{m \ \text{def.}}{>} L \overset{(3.8)}{>} S_{W_{L}}(m\!+\!m', T\!-\!1)
\] 
and since $S_{W_{L}}(\tau, T\!-\!1) < S_{W_{L}}(\tau \! - \! 1, T) $, we conclude that \textbf{the stock price cannot be a martingale.} This means that the market model admits arbitrage opportunities.\\
\vspace{0.5cm}

Let us recap and examine the assumptions (3.4)-(3.9), which led us to that conclusion. Note that all assumptions hold with sufficient discretization. It is fairly easy to describe a market model where the assumptions hold, with quite a small number of timesteps (see for example figure 3). The validity of (3.6) is of key importance and we will see how many discrete time steps it requires for the inequality to hold in case of plausible market parameters.\\

Recall that $m$ is the minimum number of upticks, for which 
\[s
\frac{X_{0}\!\cdot\!U^{m}}{N} > L  \quad \text{or equivalently} \quad m > \dfrac{\text{ln}\left(LN/X_{0}\right)}{\text{ln}(U)}. 
\]
It can be written as:
\[
m = \dfrac{\text{ln}\left(LN/X_{0}\right)}{\text{ln}(U)} + \Delta_{n}, \ \Delta_{n} \in [0, 1), \ m \in \mathbb{N}.  
\]
Using formula $\text{log}_{b} c = \dfrac{\text{log}_{a} c}{\text{log}_{a} b} $, we can write $ m = \text{log}_{U}\left( \frac{LN}{X_{0}} \right)+ \Delta_{n}$ and
\begin{equation}
X_{0}\!\cdot\!U^{m} = X_{0}\!\cdot\! \left( \frac{LN}{X_{0}}\!\cdot\!U^{\Delta_{n}} \right) = LNU^{\Delta_{n}}.
\end{equation} 
By multiplying both sides of (3.6) by $N+M$, we get
\[
\frac{N+M}{N} \cdot X_{0}U^{m} \! \cdot \! D > X_{0}U^{m} \! \cdot \! U + D_{t+1,T} \!\cdot\! MK.
\]
By making a substitution from (3.10), we get
\[
(N+M) \!\cdot\!LU^{\Delta_n} \!\cdot\! D > LNU^{\Delta_n} \!\cdot\! U + D_{t+1,T} \!\cdot\! MK,  \quad \mid \cdot U
\]
\[
(N+M) \!\cdot\!LU^{\Delta_n} > LNU^{\Delta_n} \!\cdot\! U^2 + D_{t+1,T} \!\cdot\! MKU,  \quad \mid : U^{\Delta_n}
\]
\[
(N+M) \!\cdot\!L > LN \!\cdot\! U^2 + D_{t+1,T} \!\cdot\! MKU^{1-\Delta_n},  \quad \mid : L
\]
\[
N+M > N \!\cdot\! U^2 + D_{t+1,T} \!\cdot\! \frac{K}{L} \!\cdot\! MU^{1-\Delta_n}. 
\]
Note that
\[
N \!\cdot\! U^2 + D_{t+1,T} \!\cdot\! \frac{K}{L} \!\cdot\! MU^2 > N \!\cdot\! U^2 + D_{t+1,T} \!\cdot\! \frac{K}{L} \!\cdot\! MU^{1-\Delta_n}.
\]
If following inequality holds:
\[
N+M > U^2 \!\cdot\! ( N   + D_{t+1,T} \!\cdot\! M\!\cdot\! \frac{K}{L} ) \ \Longleftrightarrow \ U^2 < \frac{N+M}{N+M\!\cdot\!\frac{K}{L} \!\cdot\! D_{t+1,T}  }, 
\]
then also (3.6) holds. For uptick ratio, we will use the expression given by Cox et al. in \cite{CoxRR1979}:
\begin{equation}
U = e^{\sigma\sqrt{T/n}},
\end{equation}
where $n$ is the number of discrete time steps. We reach the conclusion that (3.7) holds if
\begin{equation}
n > \frac{T}{\left[ \frac{\text{ln}\left( \frac{N+M}{N+M\!\cdot\!\frac{K}{L} \!\cdot\! D_{t+1,T} } \right)}{2\sigma} \right]^2}.
\end{equation}

For example, for market parameters $X(0)=1000$, $N=10$, $M=4$, $L=190$, $K=95$, $T=5$, $\sigma=0.4$, $r=0$, the inequality holds if $n > 134$. All in all, we have seen that the absence of martingale property exhibits itself with a quite small number of discrete time steps. 
\vspace{1cm}

\section{Consistent pricing of threshold warrants}

Ideally we would like to have a pricing model for threshold warrants which is consistent with arbitrage pricing principles and corresponding pricing theory of ordinary warrants developed in \cite{GalaiS1978, Hull}) and in many other papers. Unfortunately, as we have demonstrated in this paper, this is not possible since the corresponding stock price process cannot be arbitrage free. So we have to relax some of the assumptions. Let us discuss some reasonable possibilities.

One of the key assumptions of arbitrage pricing models is the following:\\

\textbf{A7.} \ At each time moment it is possible to buy and sell stocks for the same price.\\

Although this assumption does not hold in practice (the bid and ask prices are different on stock markets), the differences are usually so small that they can be ignored when developing a pricing framework for derivative instruments. In the case of active threshold warrants, the situation can be different, especially close to the expiration date of the warrants. Let's assume rational behaviour of market participants and consider a situation where the threshold condition has not been met, but the best ask price is above the threshold $L$.  Then nobody would buy the stock for a price that is (noticeably) higher that $S_{W}(t)$ (the arbitrage price of the share in the presence of ordinary warrants). At the same time, when the expiration date is close and the threshold condition has not been satisfied, then there does not seem to be any reasons for shareholders to sell their shares for less than the undiluted share price $S(t)=\frac{X(t)}{N}$ per share. Since close to the expiry $S(t)$ and $S_{W}(t)$ can be quite different, the differences between bid and ask prices can be large and trading with the shares would stop for certain market scenarios. 

If the assumption A7 is dropped, then in order to get a pricing model for threshold warrants that is consistent with arbitrage pricing of ordinary warrants, it is necessary to:
\begin{enumerate}
\item \ Choose a model for $X(t)$.
\item \ Specify models for processes $Ask(t)$ and $Bid(t)$ satisfying requirements $S_{W}(t)\leq Bid(t)\leq Ask(t)\leq S(t)$ together with probabilities $p(t)$ for a trade to happen at a price that is higher than $L$ at time $t$ so that after the threshold condition has been met, we have $Ask(t)=Bid(t)=S_{W}(t)$.
\item \ Choose a pricing principle.
\item \ Compute the price by a suitable numerical method.
\end{enumerate}
By this approach it is possible to exclude arbitrage possibilities from the market but we need to model three processes in addition to $X(t)$.

Another possibility is to keep A7 together with the assumption that at least one trade happens at every time moment, but to allow limited arbitrage opportunities for the stock price process. For this, it is necessary to:
\begin{enumerate}
\item \ Choose a model for $X(t)$.
\item \ Specify models for the process $S_{W_L}(t)$ satisfying the requirements $S_{W}(t)\leq S_{W_L}(t)\leq S(t)$.
\item \ Specify limits for the number of shares traded at each time moment.
\item \ Choose a pricing principle. 
\item \ Compute the price by a suitable numerical method.
\end{enumerate}
Since the first possibility leads to an incomplete market model and the second one to a market model with arbitrage possibilities, the arbitrage (or risk-neutral) pricing principles cannot be used to derive the fair price for threshold warrants. We propose to use the indifference pricing principle to compute the prices of threshold warrants.\\

\vspace{1cm}

\section{Indifference pricing approach}\label{s:4} 

Since the stock price models discussed above are incomplete and/or admit limited arbitrage opportunities and thus arbitrage pricing principles cannot be applied, we introduce an alternative approach. Consider two definitions of \textit{price:}\\

\textbf{Def A.} The least amount of money to cover a claim at an acceptable level of risk.\\

\textbf{Def B.} The least amount of money one could sell a claim for, without worsening one's risk profile.\\

The acceptable level of risk in definition A may also mean no risk at all. In the case of complete markets, all risk can be removed by replication. Then the least amount of money to cover a claim is simply the price of creating the replicating portfolio. Therefore, definition A is a generalisation of the arbitrage pricing approach that is used in the Black-Scholes framework. Definition B describes the reality where we do not sell/buy assets if it is not somehow beneficial to us. Here we use the latter definition\footnote{For further reading on the subject, see \cite{Carmona}.}. First we describe our risk profile by choosing a utility function. It is a nondecreasing, marginally decreasing and concave function, which describes the usefulness of a certain sum of money for us. For example, exponential utility:
\begin{equation}
u(x) = -\frac{1}{\gamma}e^{-\gamma x}, \ \gamma>0.
\end{equation}

Then we construct a self-financing portfolio for every timestep with value 
\[
 V(t,\delta) =   
  \begin{cases}
   w(t,\delta), 					&  t \leq T, \\  
   w(t,\delta) + C(S_{W_L}(t)), 	&  t = T,
  \end{cases}
\]
where $C(S_{W_L})$ is a receivable claim from derivative and $w_t$ is our wealth by time $t$, obtained by buying $\delta$ shares in the previous time point:
\[
w_t = \left[w_{t-1} - \delta\!\cdot\!S_{W_L}(t-1) \right](1+r) + \delta\!\cdot\!S_{W_L}(t).
\]
Risk free return for one time period is denoted by $r$. Our initial capital is $w_0$. By choosing $\delta$, we maximise our expected utility for every timestep $t \in \{0,\ldots,T\!-\!1\}$. The utility of an optimal trading strategy is expressed as
\[
\varphi(w_{t-1}, C) = \underset{\delta\in [a,b]}{\text{max}}\left[ E\left[ u(V(t,\delta)) \right] \right], 
\]
where $a$ and $b$ are the fraction of shares that can be bought or sold at each time step. 
Finally, \textbf{indifference price} at time $t=0$ is defined as minimal wealth $w_0$, for which
\[
\varphi(w_0, 0) \geq \varphi(0, C).
\]
It means that we are indifferent towards owning sum $w_0$ or derivative with claim $C$. For computing indifference price, dynamic programming must be used. This means that we compute $\varphi(w_{T-1}, C)$ at time $T\!-\!1$ for some set of $w_{T-1}$-s. At time $T\!-\!2$ we fix a certain initial wealth. The optimal trading strategy (i.e. choice of $\delta$) is such that it provides for time $T\!-\!1$ the wealth $w^*_{T-1}$ which allows to achieve the largest $\varphi(w^*_{T-1}, C)$. We do that for a set of $w_{T-2}$-s and reiterate the procedure at $T\!-\!3$.
By these steps we achieve the optimal utility at time $t=0$ with given initial wealth.\\

To gain some understanding of the indifference pricing method, let us examine a very simple example. We will price a European call option with $K=100$ in a 1-period model, where $r=0$. We use a stock price model presented in figure 4, where the probability of an uptick is $p$. Also, here we will not constrain $\delta$ values and use the utility function in the form $u(x) = -e^{-x}$. At time $t=0$
\begin{figure}
\begin{center}
\includegraphics[scale=0.3, trim = 20mm 30mm 0mm 25mm, clip]{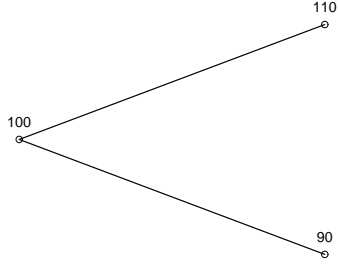} 
\caption{Simple 1-period binomial model of a stock price.}  
\end{center}
\end{figure}

\[
\varphi(0, C) = \underset{\delta\in \mathbb{R}}{\text{max}} [ -p \! \cdot \! \textrm{exp}( -\{ (0-\delta \! \cdot \! 100) \! \cdot \! (1+0) + \delta \! \cdot \! 110 +  10 \} ) -
\]
\[
 - (1-p) \! \cdot \! \textrm{exp}( -\{ (0-\delta \! \cdot \! 100) \! \cdot \! (1+0) + \delta \! \cdot \! 90 + 0 \} ) ].
\]

Here optimal $\delta = -(\textrm{ln}(1/p-1) + 10)/20$. For a $p=0.7$ we have $\delta \approx -0.458$ and $\varphi(0, C) \approx 0.006$. Without the option, but with initial wealth, we can obtain utility 
\[
\varphi(w_0, 0) = \underset{\delta\in \mathbb{R}}{\text{max}} [ -p \! \cdot \! \textrm{exp}( -\{ (w_0-\delta \! \cdot \! 100) \! \cdot \! (1+0) + \delta \! \cdot \! 110 + 0 \} ) -
\]
\[
 - (1-p) \! \cdot \! \textrm{exp}( -\{ (w_0-\delta \! \cdot \! 100) \! \cdot \! (1+0) + \delta \! \cdot \! 90 + 0 \} ) ].
\]
 
For $w_0=5$ equality $\varphi(w_0, 0) = \varphi(0, C) \approx 0.006$ holds. Since $\varphi(w_0, C)$ increases monotonically with $w_0$, our indifference price for the call option is 5. It is easy to see that we get the same answer by the replication approach. Indeed, in complete market models where trading is not constrained, the indifferent prices and prices of replicable claims coincide.\\

In multi-period setting we have to use the dynamic programming as described before. Also, for pricing threshold warrants, we have to set the trading constraints and specify a model for the process $S_{W_{L}}(t)$.
\vspace{1cm}

\section{Summary}\label{s:5} 
For pricing threshold warrants, we need a model for the stock price. By  modelling the value of the firm, we can derive bounds for the stock price, but can also demonstrate that the stock price process cannot correspond to an arbitrage free market. Therefore it is necessary to drop some assumptions that are usually made when deriving warrant prices. Thus the price of a warrant cannot be obtained by a replication argument. Instead, indifference pricing can be used. 
Suitable continuous time models for describing the stock price process in the presence of threshold warrants and methods for consistently estimating market parameters are the aim of our future research.  

\bibliography{x}
\end{document}